\documentclass[a4paper,11pt]{article}
\usepackage{pos}

\title{The $\pi^0\to \gamma^\ast \gamma^\ast$ transition form factor and the pion
  pole contribution to $a_{\mu}$ on CLS ensembles}
\ShortTitle{$\pi^0\to \gamma^\ast \gamma^\ast$ transition form factor}

\author*[a]{Jonna~Koponen}
\author[b]{Antoine G\'erardin}
\author[a,c]{Harvey~B.~Meyer}
\author[a]{Konstantin Ottnad}
\author[a]{Georg von Hippel}

\affiliation[a]{PRISMA$^+$ Cluster of Excellence \& Institut f\"{u}r Kernphysik,
  Johannes-Gutenberg-Universit\"{a}t  Mainz,\\  D-55099 Mainz, Germany}

\affiliation[b]{Aix-Marseille Universit\'e, Universit\'e de Toulon, CNRS, CPT,\\ Marseille, France}

\affiliation[c]{Helmholtz-Institut Mainz, Johannes-Gutenberg-Universit\"{a}t Mainz,\\
D-55099 Mainz, Germany}

\emailAdd{jkoponen@uni-mainz.de}

\abstract{
  We present the status of the Mainz group's lattice QCD calculation of the pion transition form
  factor $\mathcal{F}_{\pi^0\gamma^\ast\gamma^\ast}$, which describes the interaction of an on-shell
  pion with two off-shell photons. This form factor is the main ingredient in the calculation
  of the pion-pole contribution to hadronic light-by-light scattering in the muon $g-2$.

  We use the $N_f = 2 + 1$ CLS gauge ensembles, and we update our previous work by including a
  physical pion mass ensemble (E250). We compute the transition form factor in the pion rest frame
  as well as in a moving frame in order to have access to a wider range of photon virtualities.
  In addition to the quark-line connected correlator we also compute the quark-line disconnected
  diagrams that contribute to the form factor.

  In this final stage of the analysis, we combine the result on E250 with the previous work
  published in 2019 to extrapolate the form factor to the continuum and to physical quark masses.
  Testing different ans\"atze for the fit, we explore the systematic uncertainties of the
  extrapolation. The contribution from the disconnected diagrams is also scrutinized.
}

\FullConference{The 41st International Symposium on Lattice Field Theory (LATTICE2024)\\
 28 July - 3 August 2024\\
Liverpool, UK\\}

\begin{document}
\maketitle

\section{Extracting the TFF}

The transition form factor (TFF) $\mathcal{F}_{\pi^0\gamma^\ast\gamma^\ast}$ is the main ingredient in the
calculation of the pion-pole contribution to hadronic light-by-light scattering in the muon $g-2$.
It describes the interaction of an on-shell pion with two off-shell photons, and it also gives us
information about the partial decay width $\Gamma (\pi^0\to \gamma\gamma)$.

Our calculation follows very closely the Mainz group's publication~\cite{Gerardin:2019vio} (see also
\cite{Gerardin:2016cqj}). The transition form factor is extracted from matrix elements
\begin{equation}
  M_{\mu\nu}(p,q_1)=\mathit{i}\!\int\! \mathrm{d}^4x \mathrm{e}^{\mathit{i}q_1\cdot x}\langle 0 |T\{J_\mu(x)J_\nu(0)\}|\pi^0(p)\rangle 
  = \epsilon_{\mu\nu\alpha\beta}q^\alpha_1q^\beta_2\mathcal{F}_{\pi^0\gamma^\ast\gamma^\ast}(q^2_1,q^2_2),
\end{equation}
where $J_\mu$ is the electromagnetic (EM) current. Here $q_1 = (\omega_1,\vec{q}_1)$ and
$q_2 = (E_\pi-\omega_1,\vec{p}-\vec{q}_1)$ are the four-momenta associated with the two
currents, and $p$ is the four-momentum of the pion, such that $p=q_1+q_2$.
To cover a wide range of photon virtualities, we use both the rest frame of the pion, $\vec{p}=(0,0,0)$,
and a moving frame $\vec{p}=(0,0,1)$ (in units of $2\pi/L$).

\begin{figure}[b]
\includegraphics[trim={0mm 0mm 0mm 0mm},clip,width=0.995\linewidth]{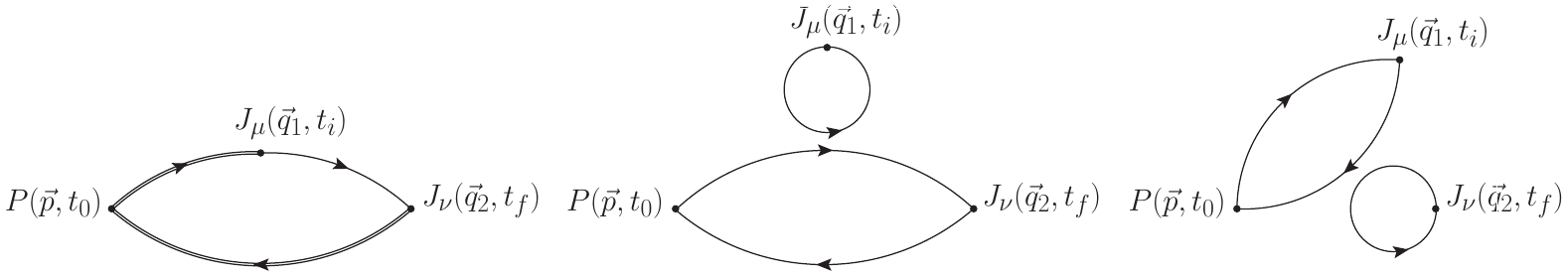}
\caption{Connected and disconnected diagrams.}
\label{fig:diagrams}  
\end{figure}  

The Euclidean matrix elements read
\begin{equation}
M_{\mu\nu} = (\mathit{i}^{n_0}) M^E_{\mu\nu},\;
M^E_{\mu\nu} = -\!\int^\infty_{-\infty}\!\mathrm{d}\tau \mathrm{e}^{\omega_1\tau}\!\int\! \mathrm{d}^3\! x \mathrm{e}^{-\mathit{i}\vec{q}_1\cdot \vec{x}}\langle 0|T\{J_\mu(\vec{x},\tau)J_\nu(\vec{0},0)\}|\pi^0(p)\rangle,
\end{equation}
where $n_0$ denotes the number of temporal indices.
These matrix elements can be obtained by integration over Euclidean time-dependent amplitude,
\begin{equation}
  M^E_{\mu\nu}(p,q_1)= \frac{2E_\pi}{Z_\pi}\!\int^\infty_{-\infty}\!\mathrm{d}\tau \mathrm{e}^{\omega_1\tau}\widetilde{A}_{\mu\nu}(\tau),
\end{equation}
where $\tau=t_i-t_f$ is the time separation between the two EM currents.
The amplitude $\widetilde{A}_{\mu\nu}(\tau)$ is connected to a 3-point correlator calculated on the lattice by
\begin{align}
  C^{(3)}_{\mu\nu}(\tau,t_\pi) &\equiv a^6\sum_{\vec{x},\vec{z}}\langle J_{\mu}(\vec{x},t_i)J_{\nu}(\vec{0},t_f)P^{\dag}(\vec{z},t_0)\rangle \mathrm{e}^{\mathit{i}\vec{p}\cdot\vec{z}}\mathrm{e}^{-\mathit{i}\vec{q}_1\cdot\vec{x}}\\
  \widetilde{A}_{\mu\nu}(\tau)&\equiv \lim_{t_\pi\to +\infty}\mathrm{e}^{E_\pi(t_f-t_0)}C^{(3)}_{\mu\nu}(\tau,t_\pi),\;\;\;\;\; (t_0 < t_f),
\end{align}
where $t_\pi$ is the time separation between the pion and the closest EM current.

For convenience we define a scalar function $\widetilde{A}^{(1)}(\tau)$ as
\begin{equation}
\widetilde{A}_{0k}(\tau) = (\vec{q}_1\times \vec{p})\widetilde{A}^{(1)}(\tau), \quad
{\epsilon'}^k \widetilde{A}_{kl}(\tau)\epsilon^l = -\mathit{i}(\vec{\epsilon}'\times \vec{\epsilon})
\cdot\left(\vec{q}_1E_\pi\widetilde{A}^{(1)}(\tau) + \vec{p}\frac{\mathrm{d}\widetilde{A}^{(1)}(\tau)}{\mathrm{d}\tau}\right).
\end{equation}
In the moving frame (at non-zero pion momentum) we also define
$\widetilde{A}_{12}(\tau) \!\equiv\! -\mathit{i}E_\pi p_z\widetilde{A}^{(2)}(\tau)$.

In addition to the quark-line connected diagram, there are contributions from two quark-line disconnected diagrams
that have to be calculated. Both the connected and disconnected diagrams are depicted in Fig.~\ref{fig:diagrams}.

\section{Lattice setup}

\begin{figure}[t!]
\centering  
\includegraphics[trim={0mm 0mm 0mm 0mm},clip,width=0.6\linewidth]{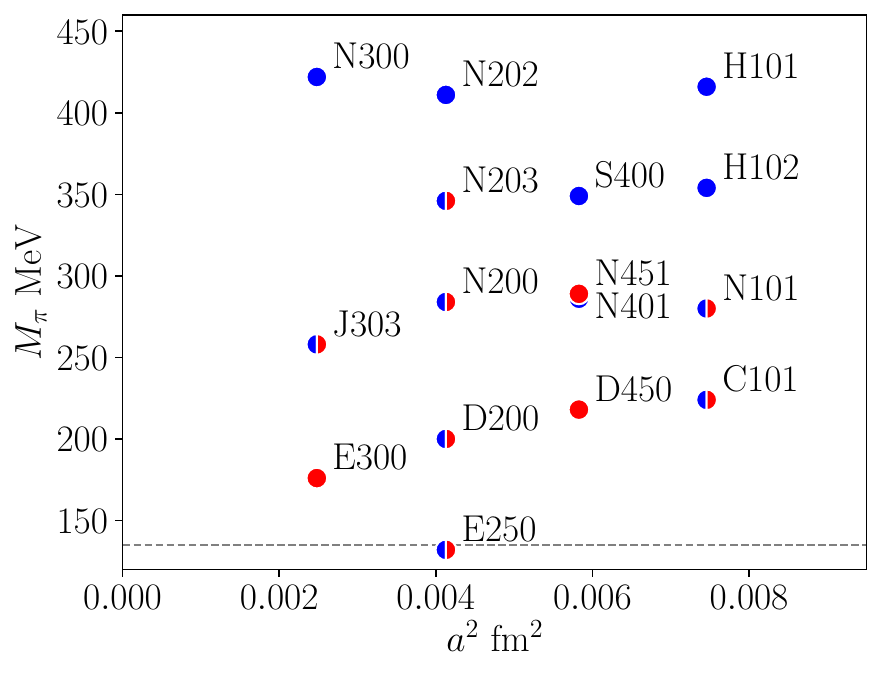}
\caption{CLS ensembles used in this study. We use data for connected pieces on ensembles marked with blue,
  and for disconnected pieces on ensembles marked with red. Note that the disconnected piece has the
  flavor structure $l-s$, and vanishes at the SU(3) symmetric point (ensembles with largest pion masses here).}
\label{fig:ensembles}  
\end{figure}

We use the CLS $N_f=2+1$ ensembles with non-perturbatively $\mathcal{O}(a)$-improved Wilson fermions and
tree-level improved L\"uscher-Weisz gauge action. We have four lattice spacings and use multiple pion masses
to control the chiral extrapolation --- see Fig.~\ref{fig:ensembles}. All ensembles have fairly large volumes
($M_{\pi}L\ge 4$). More details about the ensembles can be found in~\cite{Gerardin:2019vio} and references
therein. For the connected piece, we now add one ensemble (E250) at the physical pion mass with a lattice
spacing of $a\approx 0.064$~fm, size $96^3\times 192$ and $L\approx 6$~fm, compared to the
publication~\cite{Gerardin:2019vio}. The disconnected contributions are calculated using entirely new data.

\section{Analysis steps and preliminary results}

\subsection{Modeling the tail}

\begin{figure}
\centering  
\includegraphics[trim={0mm 0mm 0mm 0mm},clip,width=0.49\textwidth]{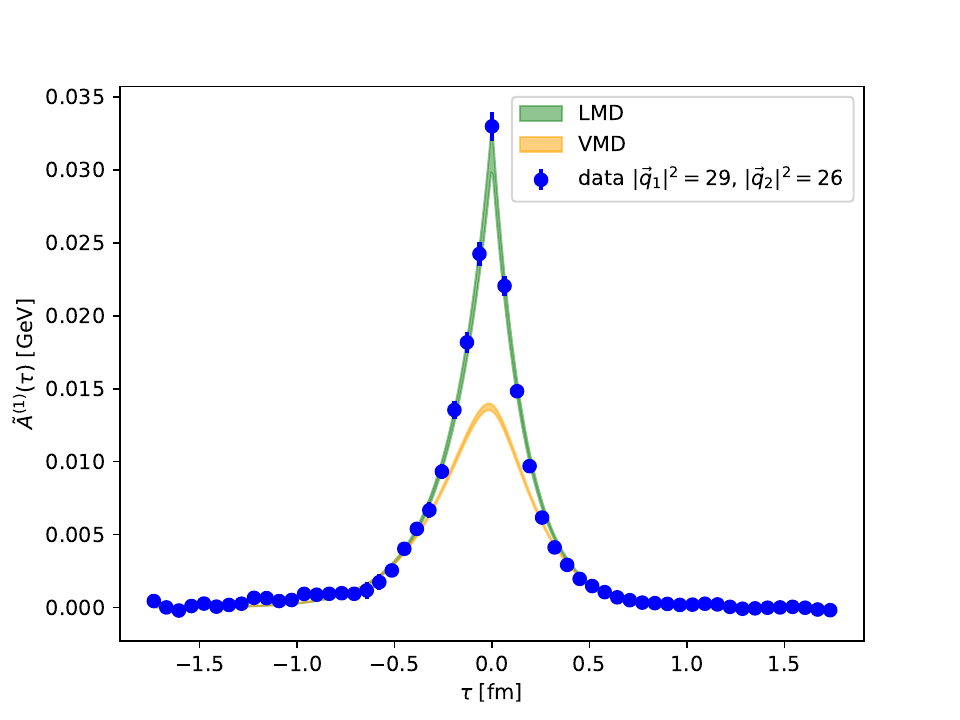}
\includegraphics[trim={0mm 0mm 0mm 0mm},clip,width=0.49\textwidth]{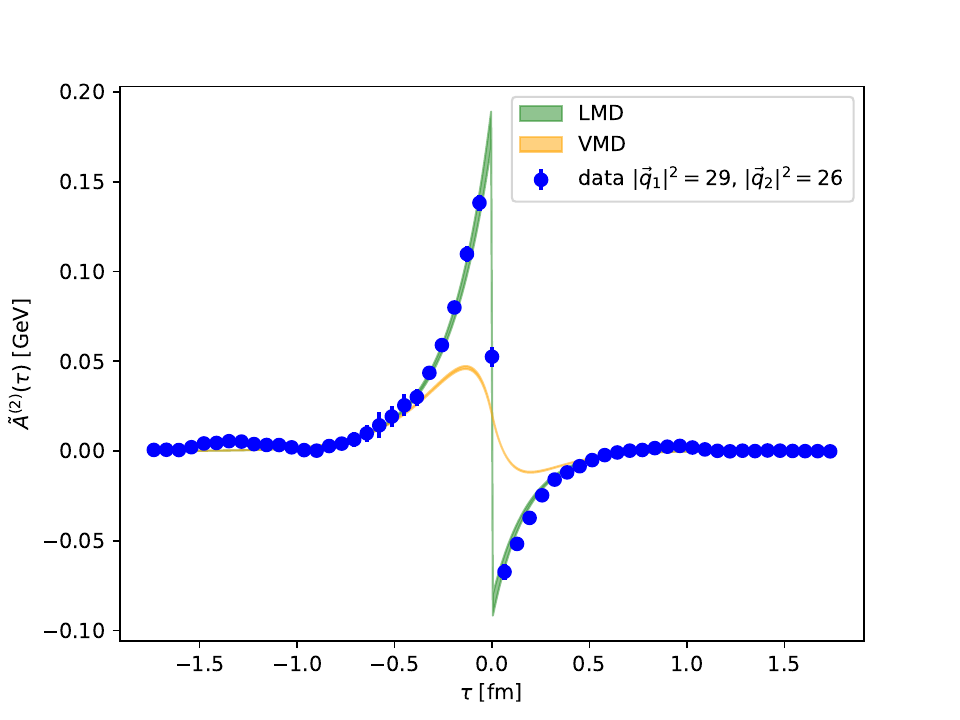}
\caption{Modeling the tail with LMD or VMD model. The tail is only used at $|\tau| > 1.2$~fm.
Lattice data  on ensemble E250 is shown here to illustrate the quality of the fit.}
\label{fig:tail}
\end{figure}

Recall that we need to integrate over $\tau$ (the distance between the two electromagnetic currents) to
extract the TFF. Therefore we need to model the tail contribution of $\widetilde{A}_{\mu\nu}(\tau)$.
For this we test both the lowest meson dominance (LMD) and the vector meson dominance (VMD) models:
\begin{equation}
    \widetilde{A}^{\textrm{LMD}}_{\mu\nu}(\tau)=\frac{Z_\pi}{4\pi E_\pi}\int_{-\infty}^{\infty}\mathrm{d}\tilde{\omega}
\dfrac{\left(P^E_{\mu\nu}\tilde{\omega}+Q^E_{\mu\nu}\right)\left(\alpha M_V^4+\beta(q_1^2+q_2^2)\right)}{\left(\tilde{\omega}-\tilde{\omega}_1^{(+)}\right)\left(\tilde{\omega}-\tilde{\omega}_1^{(-)}\right)\left(\tilde{\omega}-\tilde{\omega}_2^{(+)}\right)\left(\tilde{\omega}-\tilde{\omega}_2^{(-)}\right)}\mathrm{e}^{-\mathit{i}\tilde{\omega}\tau},
\end{equation}
with
\begin{align*}
P_{\mu\nu}^E=&\mathit{i}\epsilon_{\mu\nu0i}p^i,& \tilde{\omega}_1^{(\pm)}&=\pm\mathit{i}\sqrt{M^2_V+|\vec{q}_1|^2},\\
  Q^E_{\mu\nu}=&\epsilon_{\mu\nu i0}E_\pi q_1^i-\mathit{i}\epsilon_{\mu\nu ij}q^i_1p^j,& \tilde{\omega}_2^{(\pm)}&=-\mathit{i}\left(E_\pi \mp\sqrt{M^2_V+|\vec{q}_2|^2}\right).
\end{align*}
This gives an explicit expression for $\widetilde{A}^{\textrm{LMD}}_{\mu\nu}$, which we use to
fit our data using $\alpha$, $\beta$ and $M_V$ as fit parameters.
The VMD model is obtained by setting $\beta=0$ in the LMD model.
The model is only used for the tail $|\tau| > 1.2$~ fm. See Fig.~\ref{fig:tail} for illustration
of the two models.

\subsection{Parameterizing the TFF (connected contribution)}

\begin{figure}
\centering  
\includegraphics[trim={0mm 0mm 0mm 0mm},clip,width=0.49\textwidth]{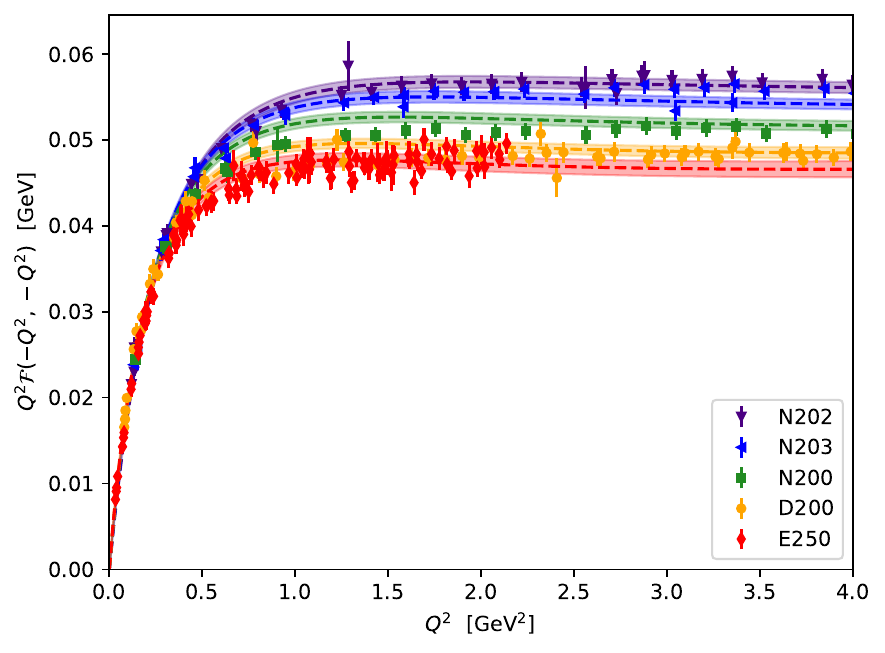}
\includegraphics[trim={0mm 0mm 0mm 0mm},clip,width=0.49\textwidth]{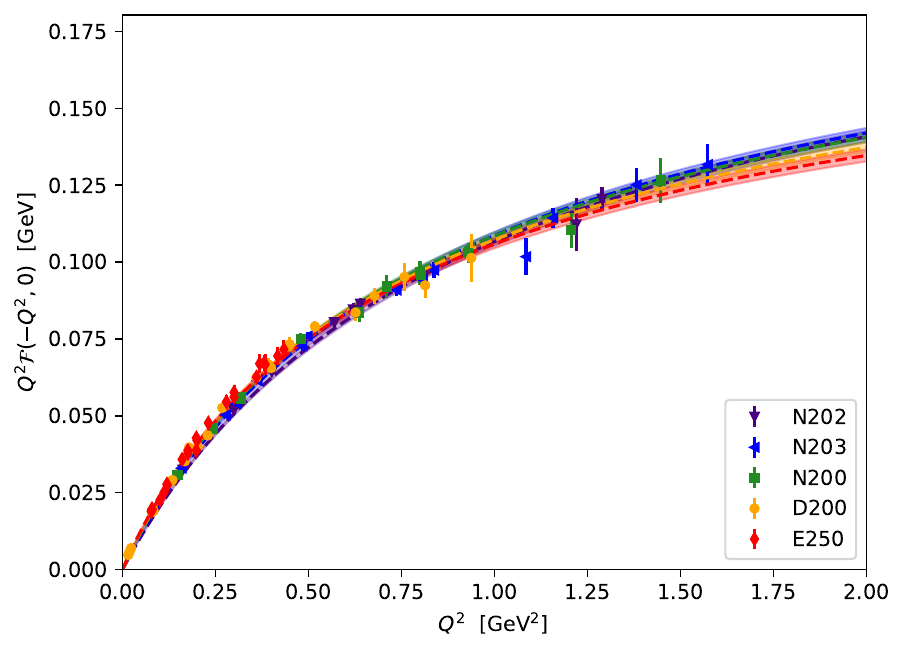}\\
\includegraphics[trim={0mm 0mm 0mm 0mm},clip,width=0.49\textwidth]{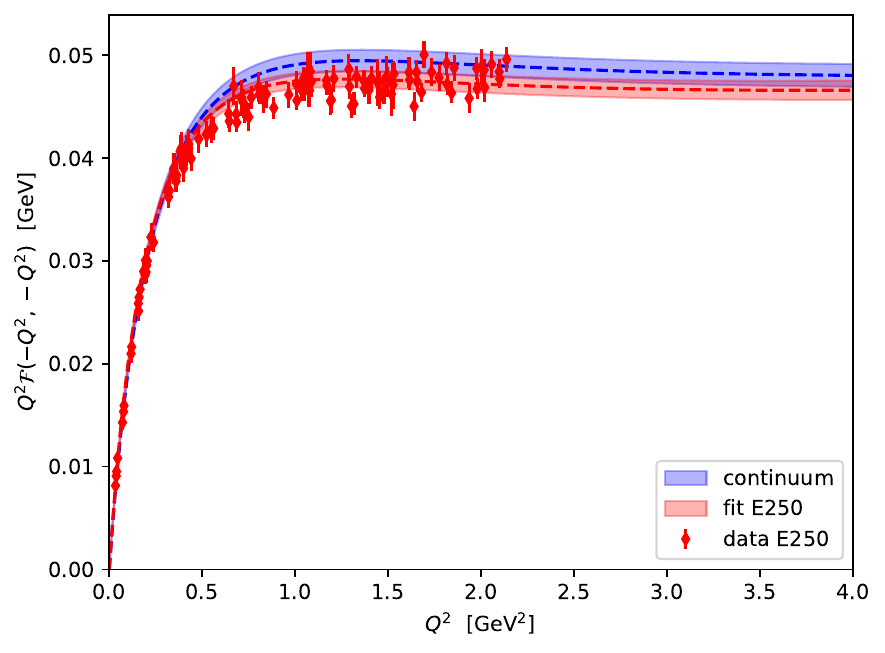}
\includegraphics[trim={0mm 0mm 0mm 0mm},clip,width=0.49\textwidth]{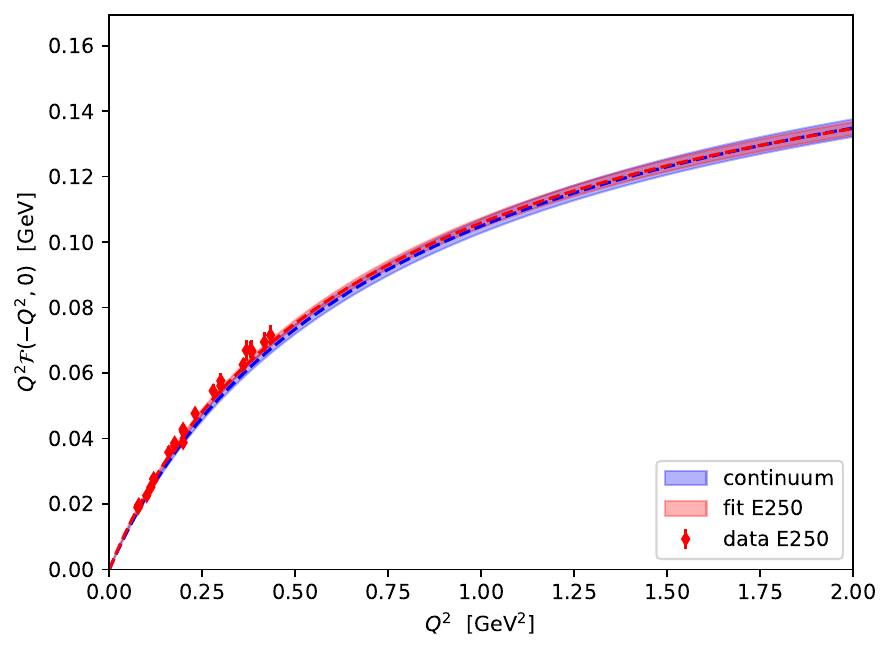}
\caption{Illustration of the fit to the TFF using the dispersion theory inspired ansatz. The ensembles shown in the top
  row all have the same lattice spacing, but different pion masses. The bottom row illustrates the magnitude of $a^2$ effects,
  comparing the continuum result with the result at E250 parameters. The pion mass on E250 is close to the physical value.
  The left column shows the 'double virtual' case $\mathcal{F}(-Q^2,-Q^2)$, whereas the right column shows the 'single virtual'
case $\mathcal{F}(-Q^2,0)$.}
\label{fig:TFFfit}
\end{figure}

Based on dispersive representation of the form factor, we start with a simplified fit ansatz
\begin{align}
    \mathcal{F}_{\pi^0\gamma^\ast\gamma^\ast}(q^2_1,q^2_2)=&\dfrac{c_1}{\left(1-q^2_1/M^2_1\right)\left(1-q^2_2/M^2_1\right)}
    +\dfrac{c_2}{\left(1-q^2_1/M^2_2\right)\left(1-q^2_2/M^2_2\right)}\\
    &+c_3q^2_1q^2_2\int_{s_{\mathrm{min}}}^\infty\dfrac{\mathrm{d}s}{\left(s-q^2_1\right)^2\left(s-q^2_2\right)^2},
\end{align}
where the masses and the coefficients $c_i$ are fit parameters. The last term guarantees
the correct asymptotic behavior at large virtualities.

This ansatz can be further improved by replacing the first term by
$\mathcal{F}_{\mathrm{vs}}(q^2_1,q^2_2)+\mathcal{F}_{\mathrm{vs}}(q^2_2,q^2_1)$,  
where the first virtuality corresponds to an isovector (v) and the second virtuality to an
isoscalar (s) photon \cite{Hoferichter:2014vra}.
At fixed isoscalar virtuality, one can write~\cite{Hoferichter:2012pm}
\begin{equation}
  \mathcal{F}_{\mathrm{vs}}(q^2_1,q^2_2)=\dfrac{1}{12\pi^2}\int_{4m_\pi^2}^\infty \!\!\mathrm{d}s
  \dfrac{q^3_\pi(s)\left(F^V_\pi(s)\right)^\ast f_1(s,q^2_2)}{\sqrt{s}\left(s-q^2_1\right)},
\end{equation}
where $s$ is the invariant mass of the $\pi^+\pi^-$ system and $q_\pi(s)=\sqrt{s/4-m_\pi^2}$,
$F^V_\pi(s)$ is the pion vector form factor,
and $f_1(s,q^2)$ is the amplitude for the process $\gamma^\ast_s\to \pi^+\pi^-\pi^0$ .
We use the Gounaris--Sakurai model \cite{Gounaris:1968mw} for $F^V_\pi(s)$, and 
\begin{equation}
f_1(s,q^2)=\left(\dfrac{c_\omega}{1-q^2/M_\omega^2}+\dfrac{c_\phi}{1-q^2/M_\phi^2}\right)\dfrac{\sqrt{s}}{q^3_\pi(s)}\mathrm{e}^{\mathit{i}\delta_1(s)}\sin \delta_1(s).
\end{equation}
Combining $F^V_\pi(s)$ and $f_1(s,q^2)$, and dropping the $c_\phi$ term, we have
\begin{equation}
    \mathcal{F}_{\mathrm{vs}}(q^2_1,q^2_2)=\dfrac{1}{12\pi^2}\left(\dfrac{c_\omega}{1-q^2_2/M_\omega^2}\right)
    \int_{4m_\pi^2}^\infty \!\!\mathrm{d}s\dfrac{q^3_\pi(s)\left|F^V_\pi(s)\right|^2}{f_0\sqrt{s}\left(s-q^2_1\right)}.
\end{equation}
This integral is done numerically.

This dispersion-theory inspired ansatz can be used to parameterize the TFF on a single ensemble, or used
as a global model to perform a chiral and continuum extrapolation. We proceed by fixing the coupling $g_{\rho\pi\pi}$
for each ensemble based on a simple fit in $m_\pi^2$ and $a^2$ on lattice data obtained in a spectroscopy
study \cite{Gerardin:2019rua} on a subset of the same CLS ensembles used in this TFF study. Once 
$g_{\rho\pi\pi}$ is fixed, we can write the $\rho$ meson width as $\Gamma_\rho=g^2_{\rho\pi\pi}k^3_\rho/(6\pi M^2_\rho)$,
where $k_\rho=\sqrt{M_\rho^2/4-M_\pi^2}$.
This leaves us six free fit parameters: $M_\omega$, $M_2$, $k_\rho$, $c_\omega$, $c_2$, and $c_3$, and their
dependence on pion mass and lattice spacing. To stabilize the fit we fix $M_2$ to be roughly 400 MeV above
the $ \rho$ mass, and set $M_\omega = M_\rho$. The fit is illustrated in Fig.~\ref{fig:TFFfit}.

\subsection{Parametrizing the disconnected contribution}

\begin{figure}
\centering  
\includegraphics[trim={0mm 0mm 0mm 0mm},clip,width=0.49\textwidth]{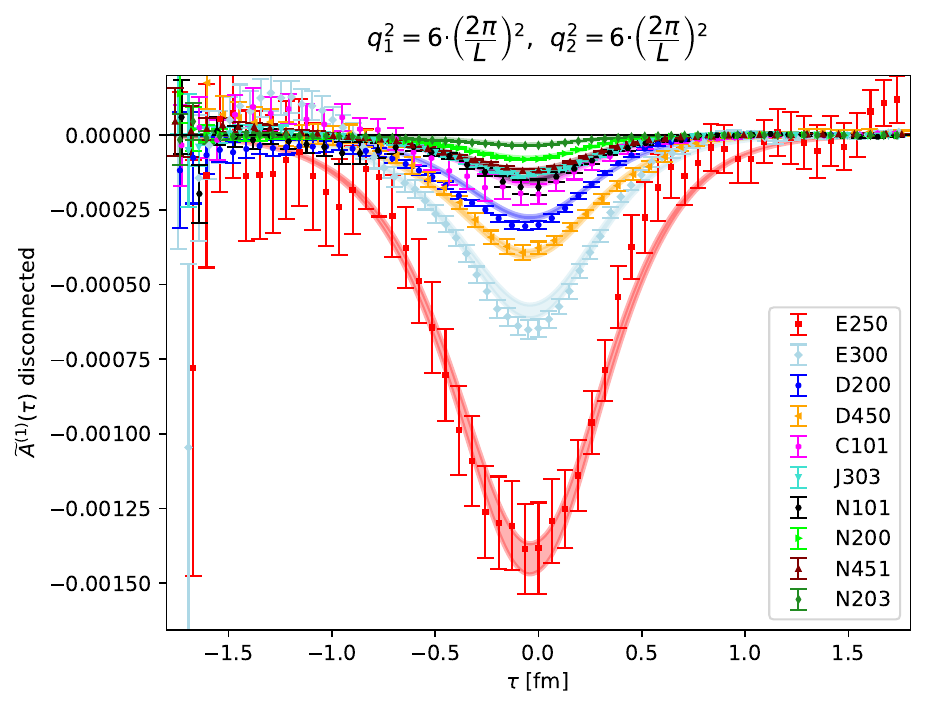}
\includegraphics[trim={0mm 0mm 0mm 0mm},clip,width=0.49\textwidth]{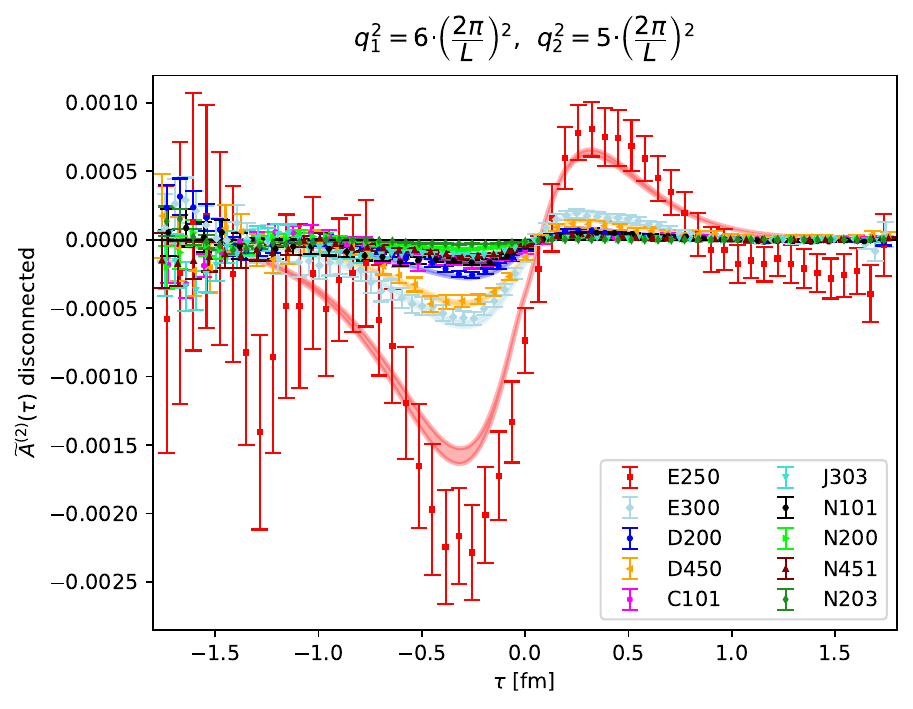}
\caption{Disconnected contribution to $\widetilde{A}^{(1)}$, $\widetilde{A}^{(2)}$: illustrating
  the fit quality across ensembles.}
\label{fig:discDVMD}
\end{figure}

We parameterize the disconnected contribution using a sum of two VMD models, i.e. we have two vector
mesons with masses $M_{V_1}$, $M_{V_2}$. The first mass is fixed to 775 MeV (the $\rho$ meson mass), and
both masses as well as the coefficients are allowed to depend on pion mass and lattice spacing. We also
include a term $m^2_\pi\mathrm{exp}(-m_\pi L)$ to take into account the finite volume of the box. This
gives a good description of data across the ensembles, which is illustrated in Fig.~\ref{fig:discDVMD}.
The disconnected contribution to the TFF is negative, but small, at most a few percent of the connected
contribution across the whole $Q^2$ range studied here.

\subsection{Pion pole contribution to $a_\mu^{\mathrm{HLbL}}$}

Once we have extracted the transition form factor for $\pi^0\to \gamma^\ast\gamma^\ast$,
we can calculate the pion-pole contribution
to hadronic light-by-light scattering, which is given by~\cite{Jegerlehner:2009ry, Nyffeler:2016gnb}
\begin{equation}  
  a_\mu^{\mathrm{HLbL};\pi^0}
  = \int_0^{\infty} \!\!\mathrm{d}Q_1\int_0^{\infty}\!\!\mathrm{d}Q_2\int_{-1}^{1}\!\!\mathrm{d}\!\cos\theta\; (F_1+F_2),
\end{equation}
with
\begin{align}  
F_1 &= f_1(Q_1,Q_2,\cos\theta)\mathcal{F}_{\pi^0\gamma^\ast\gamma^\ast}(-Q^2_1,-(Q_1+Q_2)^2)\mathcal{F}_{\pi^0\gamma^\ast\gamma^\ast}(-Q_2^2,0)\\
F_2 &= f_2(Q_1,Q_2, \cos\theta)\mathcal{F}_{\pi^0\gamma^\ast\gamma^\ast}(-Q^2_1,-Q_2^2)\mathcal{F}_{\pi^0\gamma^\ast\gamma^\ast}(-(Q_1+Q_2)^2,0).
\end{align}  
Here $\theta$ is the angle between the two momenta $Q_1$ and $Q_2$, and $f_1$ and $f_2$ are known, dimensionless
weight functions. The plan is to use the extrapolated, physical TFF to calculate $a_\mu^{\mathrm{HLbL};\pi^0}$, and
to compare that to calculating $a_\mu^{\mathrm{HLbL};\pi^0}$ on each ensemble and then extrapolating
$a_\mu^{\mathrm{HLbL};\pi^0}$ to physical pion mass and continuum.

\section{Summary and outlook}

In this report, we have given a status update of the Mainz group’s calculation of the pion
transition form factor. The main development, in addition to the inclusion of a physical pion mass ensemble
E250, is the computation of the disconnected diagrams needed in addition to the quark-line connected piece
to construct the full form factor. The results presented here are still preliminary, as we work on the final
details of the analysis.

The transition form factor $\mathcal{F}_{\pi^0\gamma^\ast\gamma^\ast}$ is the main ingredient in the estimation of
the pion-pole contribution to hadronic light-by-light scattering in the muon $g - 2$. The goal of this work is
to improve this estimate in two ways: including a physical pion mass ensemble helps constrain the chiral
extrapolation, and the precise knowledge of the disconnected contribution helps reduce the systematic error
from that source. All in all, having a precise calculation of the pion-pole contribution is a valuable
complement to direct lattice QCD calculations of the hadronic light-by-light contribution. The normalization of
the transition form factor, $\mathcal{F}_{\pi^0\gamma^\ast\gamma^\ast}(0,0)$, can also be compared to experimental
results on the neutral pion lifetime, or decay width $\Gamma(\pi^0\to\gamma\gamma)$, like PrimEx-II
\cite{PrimEx-II:2020jwd}.

\acknowledgments{
The authors acknowledge the support of Deutsche Forschungsgemeinschaft (DFG) through project
HI 2048/1-3 ``Precise treatment of quark-sea contributions in lattice QCD'' (project 399400745),
through the research unit FOR~5327 ``Photon-photon interactions in the Standard Model and beyond
-- exploiting the discovery potential from MESA to the LHC'' (grant 458854507),
and through the Cluster of Excellence ``Precision Physics, Fundamental Interactions and Structure of Matter''
(PRISMA+ EXC 2118/1) funded within the German Excellence Strategy (project ID 39083149).
We thank our colleagues in the CLS initiative for sharing ensembles.

We acknowledge PRACE for awarding us access to HAWK at GCS@HLRS, Germany via application 2020225457.
Parts of this research were conducted using the supercomputer MOGON 2 offered by Johannes Gutenberg
University Mainz (hpc.uni-mainz.de), which is a member of the AHRP (Alliance for High Performance
Computing in Rhineland Palatinate,  www.ahrp.info) and the Gauss Alliance e.V.
The authors gratefully acknowledge the Gauss Centre for Supercomputing e.V. (www.gauss-centre.eu) for
funding this project by providing computing time on the GCS Supercomputer SuperMUC-NG at Leibniz
Supercomputing Centre (www.lrz.de) and on the GCS Supercomputer JUWELS at Jülich Supercomputing Centre (JSC).
}

\end{document}